%% file: main.tex
\documentclass[aps,prd,twocolumn,superscriptaddress,nofootinbib,floatfix]{revtex4-2}

\usepackage{amsmath,amssymb,bm}
\usepackage{graphicx}
\usepackage{booktabs}
\usepackage[colorlinks=true,linkcolor=blue,citecolor=blue,urlcolor=blue]{hyperref}

\newcommand{\dd}{\mathrm{d}}
\newcommand{\eps}{\varepsilon}
\newcommand{\PB}[2]{\{#1,#2\}}
\newcommand{\kk}{\bm{k}}
\newcommand{\Ups}{\bm{\Upsilon}}
\newcommand{\kres}{\bm{\kappa}}
\newcommand{\code}[1]{\texttt{#1}}
\newcommand{\sinc}{\operatorname{sinc}}

\begin{document}

\title{Resonance crossings as entire functions of the Koopman operator}
\author{Priscilla Canizares}
\affiliation{DAMTP, Centre for Mathematical Sciences, University of Cambridge, Wilberforce Road, Cambridge CB3 0WA, UK}

\date{\today}

\begin{abstract}
Astrophysical binaries formed by a stellar-mass compact object and a
massive black hole are among the most promising sources for future
space-based gravitational-wave detectors. These extreme mass-ratio
inspiral (EMRI) systems are tracked coherently over more than $10^{5}$
orbits, encoding detailed information about strong-field gravity.
Translating a detection into physical parameters requires models of
comparable precision. One outstanding difficulty is the treatment of
transient orbital resonances. In this work we address the problem at
the level of the Koopman operator, where the waveform is promoted to
an observable and the evolution over one radial cycle is a linear
operator. Removing the fast timescale without a singularity at a
crossing is then a matter of choosing a function of that operator. We demonstrate the approach on a Kerr geodesic driven through the
$3\!:\!2$ crossing by a modelled forcing term. There the coefficient of the
standard near-identity transformation diverges, while the
finite-window generator attains its analytic bound to better than
one part in $10^{6}$. Averaging over twice the crossing
duration recovers the resonant jump with its predicted magnitude and
phase dependence.
\end{abstract}

\maketitle

\input{sections/intro}

\input{sections/methods}

\input{sections/results}
\input{sections/summary}

\section*{Data availability}
\label{data_av}
The tidal-resonance fitting formulas used in Sec.~\ref{sec:tidal} are publicly available\footnote{Black Hole Perturbation Club, \url{https://sites.google.com/view/bhpc1996/data}}. The
resonant-orbit parameters and flux variations of App.~\ref{sec:external} and
Sec.~\ref{sec:selfforce} are obtained from Ref.~\cite{FlanaganHughesRuangsri2014}. The frequency measurements, located
crossings and generated grids underlying the figures are available from the
author on request.

\begin{acknowledgments}
PC thanks Prof.~Igor Mezi\'{c} and the members of the LISA Consortium
\emph{EMRI with tidal resonances} working group for useful
discussions. PC also thanks Prof. Carola-Bibiane Sch\"onlieb for her support, and thanks the Isaac Newton Institute for
Mathematical Sciences, Cambridge, for support and hospitality during
the programme \emph{Operator methods for dynamical systems}, where
part of this work was undertaken. This work was supported by EPSRC
grant EP/Z000580/1.
\end{acknowledgments}


\appendix
\makeatletter
\def\@sectioncntformat#1{\csname the#1\endcsname.}
\def\@hangfrom@section#1#2#3{#1\@if@empty{#2}{#3}{#2\@if@empty{#3}{}{\ #3}}}
\makeatother
\section*{Appendix}

\input{sections/appendices}

\bibliographystyle{apsrev4-2}
\bibliography{refs}

\end{document}

%% file: sections/intro.tex
\section{Introduction}
\label{sec:intro}

Extreme mass-ratio inspirals (EMRIs), binaries formed by a stellar-mass compact
object and a massive black hole, are among the most promising sources for future
space-based gravitational wave detectors~\cite{Babak2017,Colpi2024}. In black
hole perturbation theory these systems can be solved to high precision:
expanding in the mass ratio, the spacetime of the massive black hole can be
treated as a fixed background and the smaller body as a perturbation of
it~\cite{BarackPound2019}.

On the orbital timescale the smaller body follows a trajectory close to a
geodesic of that background---departing from it under the effect of the
self-force it generates. These geodesics have
a radial frequency $\Omega_r$, a polar frequency $\Omega_\theta$ and an
azimuthal frequency $\Omega_\phi$~\cite{Carter1968,Schmidt2002}, all available
in closed form~\cite{Schmidt2002,FujitaHikida2009}, and in the Mino time
parameterisation the radial and polar motions decouple~\cite{Mino2003}.

The self-force has a dissipative part (radiation reaction), which
drives the inspiral, and a conservative part shifting the
short-timescale motion relative to the background
geodesic~\cite{BarackPound2019}.

Along a geodesic the motion is periodic in the orbital phases, so
each component of the self-force can be expanded as a Fourier series
with frequencies $k\Omega_\theta + n\Omega_r$~\cite{HindererFlanagan2008}.
Because the radiation-reaction timescale is long compared with the
orbital one, every term with $k\Omega_\theta + n\Omega_r \neq 0$\footnote{Notice that, since Kerr is axisymmetric and the driving force
considered carries no azimuthal dependence, the azimuthal frequency
does not appear (Sec.~\ref{sec:methods}).}
oscillates and averages to zero, so the secular evolution is carried by the $k = n = 0$
component alone. This is the adiabatic approximation that makes long inspiral evolution tractable.

However, due to the EMRI dynamics, there are particular cases where small integers $\beta_\theta$ and
$\beta_r$ make some Kerr orbits commensurate, with
$\Omega_\theta/\Omega_r = \beta_\theta/\beta_r$. On such an orbit there exist
nonzero $(k,n)$ for which
\begin{equation}
  k\Omega_\theta + n\Omega_r = 0,
  \label{eq:resonance}
\end{equation}
and the corresponding Fourier terms no longer oscillate---surviving the average and contributing to the secular evolution in the same way
as the $k = n = 0$ term. Since the frequency ratio evolves monotonically as the
orbit shrinks, each such resonance is crossed
once~\cite{HindererFlanagan2008,FlanaganHinderer2012}.

Every inspiral encounters at least one low-order resonance, and each
crossing lasts hundreds of orbital cycles at mass ratio
$10^{-6}$~\cite{RuangsriHughes2014}; the few-cycle dephasings induced
by $3\!:\!2$ crossings alone can bias parameter estimates by several
times the typical measurement precision~\cite{SperiGair2021}. For LISA, where an EMRI is
tracked coherently over $10^{5}$ or more cycles, a mismodelled resonance
crossing can introduce systematic errors that propagate into every inferred
parameter~\cite{Berry_2016}. 

To evolve the EMRI dynamics beyond the adiabatic approximation, the
near-identity scheme removes the orbital timescale by absorbing the
oscillating part of the forcing into new
variables~\cite{Deprit1969,Brouwer1959,%
vandeMeentWarburton2018}, and its own statement of validity is that it applies
away from orbital resonances; the transformation is defined by a cohomological
equation whose solution carries $\kk\cdot\bm{\Omega}$ in the denominator, where
$\kk=(k,n)$ collects the integer harmonic labels and
$\bm{\Omega}=(\Omega_r,\Omega_\theta)$ the fundamental frequencies they multiply. At a resonance Eq.~(\ref{eq:resonance}) this approach is, then, 
singular by construction. Lynch \emph{et al.}~\cite{Lynch2024} close the gap with a partial
averaging transformation in a neighbourhood of the crossing, together with a
criterion for where to switch between the two schemes.

The dissipative dynamics is also what drives the inspiral through its
resonances. Two recent constructions approach it in the \emph{in--in}
formalism, which makes a variational principle available for
non-conservative motion~\cite{Galley2013}. Aykroyd \emph{et
al.}~\cite{Aykroyd2025} build a canonical near-identity transformation by Lie series, obtaining closed-form generator coefficients through
2.5PN at arbitrary eccentricity, and Blanco~\cite{Blanco2026} builds
the generator as a finite-time Magnus expansion. In the operator formulation presented here, the latter appears as a particular case,
the first Magnus term of $\log U_\Lambda$ (Sec.~\ref{sec:methods}).

In our approach, the generator depends on the
detuning through a function with no pole, so no resonance need be
located in advance, no second transformation constructed, and no
switching criterion invoked. In this practical setting, one
generator covers the entire inspiral, resonances included, so a
pipeline built on it carries no crossing catalogue and nothing to
tune. The quantities to be computed are then functions of a single operator. 

Our framework is the same incremented action-angle variables as
Refs.~\cite{Galley2013,Aykroyd2025,Blanco2026}, but our construction
is at the level of the operator rather than of a particular normal
form; it is the dissipative counterpart of a structure-preserving
treatment in which a symplectic neural flow map represents the
dynamics~\cite{Canizares2024}.

In this way, the evolution over one radial period is a linear
operator on phase-space functions, $U_\Lambda$. The spectrum of this
Koopman operator~\cite{Koopman1931,Mezic2005} on the invariant torus
is $\{e^{\mathrm{i}\kk\cdot\Ups\Lambda}\}$, so in this approach a
resonance occurs where an eigenvalue reaches unity.

The operator itself is not new to the literature. It was introduced by
Koopman for Hamiltonian flows~\cite{Koopman1931} and developed into a practical
tool for nonlinear systems by Mezi\'c~\cite{Mezic2005}, and it is now widely
used in fluid dynamics, in the study of power-system stability and in
data-driven modelling generally~\cite{Schmid2010,SusukiMezic2014} (see also
Refs.~\cite{Budisic2012,Brunton2022}). In almost all of that work it is a
route to a spectral decomposition estimated from trajectories, whereas here we
neither estimate $U_\Lambda$ nor decompose it into modes, but ask
which function of it an averaging scheme should apply.

The paper is organised as follows. Section~\ref{sec:methods} develops
the formulation and describes our implementation.
Section~\ref{sec:results} reports the Kerr crossing study. We
finish in Sec.~\ref{sec:summary} with a summary and future work.

Throughout, $G = c = 1$, the central mass is $M$, and an overdot denotes a
derivative with respect to Mino time $\lambda$ except where stated.

%% file: sections/methods.tex
\section{Methods}
\label{sec:methods}

\subsection{The dynamics in action--angle variables}
\label{sec:formulation}
\label{sec:mechanism}
\label{sec:action-angle}

The binary systems we consider have bound trajectories that are
\emph{quasi-periodic}, that is confined to an invariant torus, which they wind
around at a fixed set of frequencies.  The Liouville--Arnol'd theorem supplies a canonical transformation
to action--angle variables $(J,\bm{q})$ adapted to that
structure~\cite{Arnold1989}: 
\begin{equation}
  q_i(\lambda) \;=\; \Upsilon_i\,\lambda + q_i(0) ,
  \qquad
  \Ups(J) \;\equiv\; \frac{\partial H}{\partial J} .
  \label{eq:angles}
\end{equation}
The actions $J$ label the torus and are constants of the unperturbed
motion. The conjugate angles $\bm{q}$ advance linearly, and the
fundamental frequencies $\Ups = (\Upsilon_1,\dots,\Upsilon_N)$ are
fixed by the actions alone. In this representation the separation of
timescales is explicit in the variables, and the dynamics can be
modelled by averaging.

Bound geodesics in a Kerr background
are integrable~\cite{Carter1968} and carry three actions whose fundamental
frequencies are known
analytically~\cite{Schmidt2002,FujitaHikida2009}; their angles advance
exactly linearly only in Mino time $\lambda$, in which the radial and
polar motions decouple~\cite{Mino2003}.

On the orbital timescale the trajectory still moves on
a torus of the unperturbed problem, while the torus itself drifts
slowly under the small radiation-reaction force. Written as a
dissipative perturbation of strength $\eps$, and separating the
Fourier mode that does not oscillate from those that do, the drift is
\begin{equation}
  \frac{\dd J}{\dd\lambda}
  \;=\; \eps\,\mathcal{G}_{\bm 0}(J)
      \;+\; \eps \sum_{\kk \neq \bm 0} \mathcal{G}_{\kk}(J)\,
            e^{\mathrm{i}\kk\cdot\bm{q}} .
  \label{eq:forcing}
\end{equation}
The first term is the secular drive, which survives any average over the fast
angles and carries the inspiral; the second is the oscillating remainder, which
averages to zero over an unperturbed torus.

\subsubsection{The near-identity transformation}

Let $\Lambda$ be the period of the fast motion used as the averaging window,
and write
\begin{equation}
  \delta \;\equiv\; \kk\cdot\Ups
  \label{eq:detuning}
\end{equation}
for the detuning of mode $\kk$, the rate at which that mode's phase
$\kk\cdot\bm{q}$ advances. A near-identity transformation
(NIT)~\cite{Deprit1969,Brouwer1959,vandeMeentWarburton2018} seeks new variables
$\tilde J = J + \eps\, w(J,\bm q)$ in which the oscillating part of
Eq.~\eqref{eq:forcing} is absent. Matching order by order gives the
cohomological equation $\Ups\cdot\partial_{\bm q} w = -\mathcal{G}$ away from
$\kk = \bm 0$, hence:
\begin{equation}
  w_{\kk} \;=\; \frac{\mathcal{G}_{\kk}}{\mathrm{i}\,\kk\cdot\Ups}
          \;=\; \frac{\mathcal{G}_{\kk}}{\mathrm{i}\,\delta} .
  \label{eq:nit}
\end{equation}
The denominator vanishes exactly at a resonance, and no rearrangement of the
same transformation removes it. In Sec.~\ref{sec:koopman} we show that this denominator originates in
an operation on the one-period evolution operator.

Notice that whether Eq.~\eqref{eq:nit} can be made singular at all
is a property of the system. The planar post-Newtonian binary, for
example, carries no resonance at any order, see App.~\ref{app:pn}.

\begin{figure}[!t]
\includegraphics[width=\columnwidth]{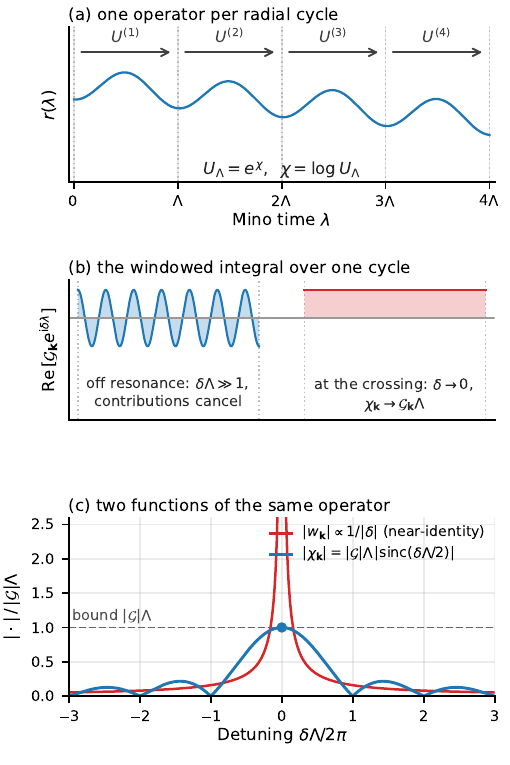}
\caption{Depiction of the finite-window construction. (a)~The inspiral is advanced
one radial cycle at a time, each cycle by the operator
$U_\Lambda = e^{\chi}$ with $\chi = \log U_\Lambda$
[Eq.~\eqref{eq:onecycle}]. (b)~Within one cycle, the windowed
integral of a mode [Eq.~\eqref{eq:magnus}]. Off resonance the
integrand oscillates and its contributions cancel; at the crossing the
phase is stationary and the integral accumulates to
$\mathcal{G}_{\kk}\Lambda$. (c)~The two functions of the same
operator, in units of $|\mathcal{G}|\Lambda$. The near-identity
coefficient applies the operator $(1-U_\Lambda)^{-1}$ and
diverges at $\delta = 0$. The
finite-window coefficient follows $|\sinc(\delta\Lambda/2)|$; it is
bounded at every detuning and saturates its bound at the crossing
[Eq.~\eqref{eq:bound}].}
\label{fig:method-schematic}
\end{figure}

\subsection{Koopman operator: the waveform as an observable}
\label{sec:observable}

A gravitational-wave (GW) model is usually described as a map from source
parameters to a time series. In the formulation used here the complex strain
$h = h_+ - \mathrm{i}h_\times$ is not a point of the binary's phase space but the
value of a \emph{function on} that space, evaluated along a trajectory. Writing
$g$ for that function, at leading quadrupole order, the waveform is
\begin{equation}
  h(t) \;=\; g\big(\phi_t(z_0)\big) \;=\; \big(U_t\,g\big)(z_0),
  \qquad U_t\,g \;\equiv\; g\circ\phi_t ,
  \label{eq:waveform-observable}
\end{equation}
with $\phi_t$ the flow and $U_t$ the Koopman operator~\cite{Koopman1931,%
Mezic2005}. Its restriction to one radial period, $t = \Lambda$, is written
$U_\Lambda$.
The trade behind Eq.~\eqref{eq:waveform-observable} is
dimensionality for linearity. The flow $\phi_t$ is a nonlinear map
on the finite-dimensional phase space. The operator $U_t$ is linear,
and it acts on the infinite-dimensional space of functions on that
phase space. Applying $U_t$ to the
coordinate functions themselves recovers the trajectory.

In the context of waveform modelling,  the objects a
waveform model manipulates are already functions on phase space, and
their time dependence is $U_t$ acting on them. On a quasi-periodic
orbit the eigenfunctions of $U_t$ are the phase harmonics
$e^{\mathrm{i}(m q_\phi + n q_r)}$, with eigenvalues
$e^{\mathrm{i}(m\Omega_\phi + n\Omega_r)t}$. Hence, when an eccentric
model is written as a sum over harmonics of
$m\Omega_\phi + n\Omega_r$, that sum is not a choice of basis. It is
the spectral expansion of $U_t$~\cite{Mezic2005,Budisic2012}. Notice that a
model that tracks amplitudes mode by mode is manipulating the
spectral data of an operator.

An eigenfunction of $U_t$ at a given frequency exists if and only if the motion
reduces, in some coordinate, to uniform rotation at that frequency. A uniformly
advancing orbital phase is therefore available only where the spectrum of $U_t$
is discrete, which holds on the quasi-periodic orbits of the adiabatic inspiral
but not as the orbit approaches plunge---there the spectrum acquires a
continuous part, and there is no phase to define.

\subsubsection{Functions of the one-cycle operator}
\label{sec:koopman}

The operators $U_t$ of Eq.~\eqref{eq:waveform-observable} form a
one-parameter family:

\begin{equation}
U_{t+t'} = U_t\,U_{t'}
\end{equation}

so the whole evolution is generated by any one member. For a radial averaging scheme, the
one-cycle operator $U_\Lambda \equiv U_{t=\Lambda}$,advances
observables by one radial period~\cite{Koopman1931,Mezic2005}:
\begin{equation}
  (U_\Lambda f)(z) \;=\; f\big(\phi_\Lambda(z)\big).
  \label{eq:koopman}
\end{equation}

$U_\Lambda$ is linear on observables however
nonlinear $\phi_\Lambda$ is, and on the invariant torus its spectrum is
$\{e^{\mathrm{i}\kk\cdot\Ups\Lambda}\}$. Therefore, a resonance occurs when an
eigenvalue reaches unity, that is, when $\delta = \kk\cdot\Ups = 0$
and hence $e^{\mathrm{i}\delta\Lambda} = 1$ (App.~\ref{app:koopman}).

The near-identity transformation removes the oscillating forcing by
diagonalising $U_\Lambda$, and doing so requires inverting
$1-U_\Lambda$. The coefficient of Eq.~\eqref{eq:nit} is
$(1-U_\Lambda)^{-1}$ applied to the forcing. An inverse of this kind 
carries a pole wherever $U_\Lambda$ has an eigenvalue equal to unity,
that is, at a resonance. In this approach the small denominator is not an artefact of how the perturbation series was
organised, but it belongs to the function being applied.

The Koopman finite-window construction composes the evolution over
$N$ radial cycles, in the spirit of stroboscopic
averaging~\cite{CalvoEtAl2011}. The result is the ordered product
$U^{(N)}\cdots U^{(1)}$ of one-cycle operators, and each advances an
observable by the Poisson action of a generator,
\begin{equation}
  \mathcal{O}_{\rm f} \;=\; e^{\PB{\chi}{\,\cdot\,}}\,\mathcal{O}_{\rm i},
  \qquad \chi \;=\; \log U_\Lambda .
  \label{eq:onecycle}
\end{equation}

The logarithm is evaluated by a windowed integral whose value at a
commensurability is finite. Truncated at its first Magnus term, this
evaluation recovers the finite-time Magnusian scheme of
Ref.~\cite{Blanco2026}, which is thus a particular case of the
operator construction. Figure~\ref{fig:method-schematic} summarises
this approach.

\begin{figure}[!t]
\includegraphics[width=\columnwidth]{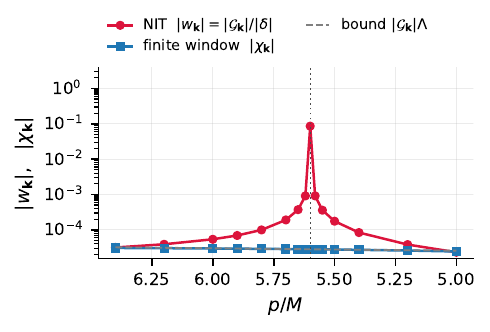}
\caption{The NIT and Koopman generators across the $3\!:\!2$ crossing at $p = 5.600\,M$, where 
p is the semi-latus rectum. The near-identity coefficient $|w_{\kk}|$ and the finite-window coefficient $|\chi_{\kk}|$  are shown against $p$, with the bound
$|\mathcal{G}_{\kk}|\Lambda$ a dashed line; the two agree
far from the crossing and separate by three orders of magnitude at it.
Our method is agnostic to the location of the crossing.}
\label{fig:boundedness-crossing}
\end{figure}

\begin{figure}[!t]
\includegraphics[width=\columnwidth]{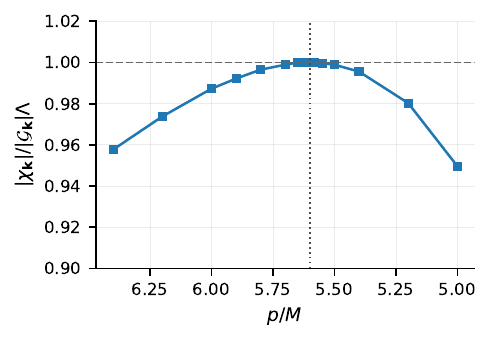}
\caption{The ratio $|\chi_{\kk}|/(|\mathcal{G}_{\kk}|\Lambda)$
across the crossing. It reaches unity at the crossing, where the bound
of Eq.~\eqref{eq:bound} is attained.}
\label{fig:boundedness-bound}
\end{figure}

In the Koopman framework, resonance crossings are treated through
\emph{entire} functions of the operator. A function of
$U_\Lambda$ acts through its values on the spectrum
$e^{\mathrm{i}\delta\Lambda}$. The near-identity route applies the operator $(1-U_\Lambda)^{-1}$.
This function is meromorphic, with poles exactly where the spectrum
reaches unity, so a crossing is a singularity of the function being
applied. The finite-window route
applies $\log U_\Lambda$, evaluated order by order in $\eps$ as
windowed integrals (Secs.~\ref{sec:first-order}
and~\ref{sec:all-orders}).

\subsubsection{First order contributions: $\chi_{(1)}$}
\label{sec:first-order}

The Magnus series~\cite{Magnus1954,Blanes2009} enters as the means of computing
$\log U_\Lambda$ for a given system---as an implementation of
Eq.~\eqref{eq:onecycle}. The generator carries two expansions at
once, in orders and in modes,
\begin{equation}
  \chi \;=\; \sum_{n \ge 1} \eps^{n}\, \chi_{(n)},
  \qquad
  \chi_{(1)} \;=\; \sum_{\kk} \chi_{\kk}\,
    e^{\mathrm{i}\kk\cdot\bm{q}_0} \;,
  \label{eq:orders-modes}
\end{equation}
where the label $(n)$ counts powers of $\eps$ in the Magnus expansion.
The subscript $\kk$ labels the Fourier modes within one order, here
the first, with $\bm{q}_0 = (q_{0,r},\, q_{0,\theta})$ the angles
at the start of the window [Eq.~\eqref{eq:pullback}]. At first order in $\eps$ the generator is the windowed
integral of the forcing, and mode by mode
\begin{align}
  \chi_{\kk}
  &= \int_0^{\Lambda} \mathcal{G}_{\kk}\, e^{\mathrm{i}\delta\lambda}\,\dd\lambda
   \;=\; \mathcal{G}_{\kk}\,
     \frac{e^{\mathrm{i}\delta\Lambda} - 1}{\mathrm{i}\delta}
  \nonumber\\
  &= \mathcal{G}_{\kk}\,\Lambda\,e^{\mathrm{i}\delta\Lambda/2}\,
     \sinc\!\Big(\frac{\delta\Lambda}{2}\Big),
  \label{eq:magnus}
\end{align}
with $\sinc(x) = \sin x/x$. 

Notice that the detuning enters Eq.~\eqref{eq:magnus} only through
the factor $\Lambda\,e^{\mathrm{i}\delta\Lambda/2}\sinc(\delta\Lambda/2)$,
which is finite for every $\delta$, in contrast with the $1/\delta$
of Eq.~\eqref{eq:nit}. The crossing is a point like any other and
needs no special treatment.

Since $|\sinc| \le 1$, the coefficients obey
\begin{equation}
  \big|\chi_{\kk}\big| \;\le\; \big|\mathcal{G}_{\kk}\big|\,\Lambda
  \label{eq:bound}
\end{equation}
for every $\kk$, including the resonant one. The bound is uniform in
the detuning, so it does not degrade as a resonance is approached. At
the crossing it \emph{saturates}. Taking $\delta \to 0$ in
Eq.~\eqref{eq:magnus} gives $\chi_{\kk} \to \mathcal{G}_{\kk}\Lambda$,
and the inequality becomes an equality. That such finite-time
integrals remain finite at resonances was observed by
Blanco~\cite{Blanco2026} for the first-order dissipative case
$\chi_{(1)}$.

The factor $\mathcal{G}_{\kk}/\mathrm{i}\delta$ is the near-identity
coefficient of Eq.~\eqref{eq:nit} itself, so that
\begin{equation}
  \chi_{\kk} \;=\; w_{\kk}\,\big(e^{\mathrm{i}\delta\Lambda} - 1\big),
  \label{eq:relation}
\end{equation}
i.e.\ the finite-window coefficient is the near-identity coefficient
times a factor that vanishes exactly where the near-identity
denominator does. This holds for every mode and every window length.
The pole of $w_{\kk}$ at $\delta = 0$ cancels against that factor,
the product stays regular, and a finite window can still capture the
kick (Fig.~\ref{fig:method-schematic}).

The same structure recurs at every order.
Section~\ref{sec:all-orders} shows that
$\chi_{(2)}, \chi_{(3)}, \dots$ involve this window factor evaluated
at combinations of detunings, together with finite differences of it,
and that there is no detuning appearing in a denominator.
\subsubsection{Second order contributions: $\chi_{(2)}$}
\label{sec:all-orders}

At second order an apparent pole of a new kind appears, and it is not
obvious whether the cancellation in Eq.~\eqref{eq:relation} survives
beyond first order. The denominators now involve the combination
$\delta_1 + \delta_2 = (\kk_1 + \kk_2)\cdot\Ups$, which vanishes
identically whenever $\kk_2 = -\kk_1$. Such a pair is exactly resonant
on every orbit. These pairs belong to the secular part of
$\chi_{(2)}$, the part that drives the inspiral, and the first-order
argument is blind to them.

Writing Eq.~\eqref{eq:forcing} in the form $\dd J/\dd\lambda = \eps\,F(J,\bm q)$, we get
\begin{equation}
  F(J,\bm q) \;=\; \mathcal{G}_{\bm 0}(J)
    \;+\; \sum_{\kk\neq\bm 0}\mathcal{G}_{\kk}(J)\,e^{\mathrm{i}\kk\cdot\bm{q}} ,
  \label{eq:field}
\end{equation}
which is the forcing expressed as a field on phase space. Evaluating it along the unperturbed
motion, with the angles replaced by their free advance
$\bm q = \Ups s + \bm q_0$, leaves a quasi-periodic function of the window
variable $s$ alone,
\begin{equation}
  \tilde F(s) \;=\; F\big(J,\,\Ups s + \bm q_0\big)
              \;=\; \sum_{\kk} \mathcal{G}_{\kk}\,
                    e^{\mathrm{i}\kk\cdot(\Ups s + \bm q_0)} .
  \label{eq:pullback}
\end{equation}


The $n$th Magnus coefficient is an $n$-fold time-ordered integral
of nested brackets of $\tilde F$ over  
$\Lambda \geq s_1 \geq \cdots \geq s_n \geq 0$. Nested brackets of quasi-periodic fields are again quasi-periodic~\cite{ChartierMuruaSanzSerna2012}, so every term of $\chi_{(n)}$ is an
integral of the form
\begin{equation}
  \int_0^\Lambda\!\!\dd s_1 \int_0^{s_1}\!\!\dd s_2 \cdots
  \int_0^{s_{n-1}}\!\!\dd s_n\;
  \prod_{j=1}^{n} e^{\mathrm{i}\delta_j s_j},
  \qquad \delta_j = \kk_j\cdot\Ups .
  \label{eq:simplex}
\end{equation}
The integrand is entire in each $\delta_j$ and the domain is compact, so
the Magnus coefficient of a finite window has no pole at any order. Its
$n = 1$ case is the window integral of Eq.~\eqref{eq:magnus}, which we write
$\mathcal{W}(\delta) \equiv \int_0^\Lambda e^{\mathrm{i}\delta s}\,\dd s
= \Lambda\, e^{\mathrm{i}\delta\Lambda/2}\sinc(\delta\Lambda/2)$. This
is an entire function, with value $\Lambda$ at $\delta = 0$.

The $n = 2$ case shows how the apparent poles disappear in practice. Carrying
out the inner integral,
\begin{widetext}
\begin{equation}
  \int_0^\Lambda\!\!\dd s_1\!\int_0^{s_1}\!\!\dd s_2\,
  e^{\mathrm{i}\delta_1 s_1} e^{\mathrm{i}\delta_2 s_2}
  = \frac{\mathcal{W}(\delta_1+\delta_2) - \mathcal{W}(\delta_1)}{\mathrm{i}\,\delta_2} .
  \label{eq:simplex2}
\end{equation}
\end{widetext}
The double integral is therefore a difference quotient of the single-window
function $\mathcal{W}$. Because $\mathcal{W}$ is entire, the apparent poles at $\delta_1 = 0$ and
at the combination $\delta_1 + \delta_2 = 0$ are absent from the numerator;
and because the numerator vanishes as $\delta_2 \to 0$ at the same rate as
the denominator, the quotient tends there to the derivative
$-\mathrm{i}\,\mathcal{W}'(\delta_1)$ and is finite as well. A difference quotient of
an entire function is entire, so the second-order integral carries no
singularity in either detuning, including resonances; the
situation is the same as at first order, where the window integral appears to
diverge at $\delta = 0$ when written as
$(e^{\mathrm{i}\delta\Lambda}-1)/\mathrm{i}\delta$ and is manifestly
finite in its $\sinc$ form.

From Eq.~\eqref{eq:simplex} we can see that the detunings
$\delta_j = \kk_j\cdot\Ups$ of Eq.~\eqref{eq:detuning} appear only in
the phases
$e^{\mathrm{i}\delta_j s_j}$, whose modulus is one. Bounding each by unity gives
\begin{equation}
  \big|\chi_{(n)}\big| \;\leq\;
  c_n\big(\,\|\tilde F\|_\Lambda\,\Lambda\big)^{n},
  \qquad
  \|\tilde F\|_\Lambda \;\equiv\; \sup_{0 \leq s \leq \Lambda} \|\tilde F(s)\| ,
  \label{eq:bound-n}
\end{equation}
with $c_n$ a combinatorial constant of the Magnus expansion and $\|\cdot\|$ any
norm submultiplicative under the bracket, its constants absorbed into
$c_n$. Consequently, at any order $n$, the terms of the series do not degrade as a crossing is
approached, Eq.~\eqref{eq:bound} is the $n = 1$
case.
A bound on every \emph{coefficient} does not, however, imply
convergence of the \emph{series}. The standard sufficient condition
$\int_0^\Lambda\!\|\cdot\|\,\dd s < \pi$~\cite{Blanes2009} is a
requirement on the perturbation strength, not on the detuning.

%% file: sections/results.tex
\section{Transient resonance crossing in Kerr spacetime}
\label{sec:results}
In this section we demonstrate our approach on Kerr geodesics.  Throughout this section the
background is a Kerr black hole of spin $a = 0.9M$, and the orbit is
a generic bound geodesic of eccentricity $e = 0.2$ and minimum polar
angle $\theta_{\min} = 1.1$~rad, corresponding to an inclination
$x \equiv \cos I = 0.891$, whose semi-latus rectum $p$ evolves
through the crossing. Unless stated otherwise the driving is the
modulated drag force $f_a = -\alpha[1 + \beta\cos(2\theta)]\,p_a$,
with $\alpha = 10^{-3}$ and $\beta = 0.5$, see 
App.~\ref{sec:silent-mode}  for details, so that it carries nonzero amplitude at
the resonant mode. We also test our method using a tidal driving 
Refs.~\cite{GuptaEtAl2021,GuptaEtAl2022} in Sec.~\ref{sec:offres}. 

The geodesics are integrated numerically and we compute the corresponding
fundamental frequencies, $\Upsilon_r$ and $\Upsilon_\theta$, from the
Mino-time periods of the radial and polar motions
(App.~\ref{app:kerr-validation}). Each crossing is then located by
bisection on the detuning $\delta = \kk\cdot\Ups$,
Eq.~\eqref{eq:detuning}. In order to obtain the mode amplitudes
$\mathcal{G}_{\kk}$, the Fourier coefficients of the forcing term, we
use the projection onto the torus derived in App.~\ref{sec:torus}.
Since the drift through the crossing is prescribed, the perturbation enters a Kerr result only
through the amplitude it carries at the resonant index, see Sec.~\ref{sec:kick} for details. Finally, the
finite-window coefficients $\chi_{\kk}$ are
obtained along the evolution by integrating each mode of the forcing term
over one radial period $\Lambda$. The near-identity coefficient
$w_{\kk}$ is evaluated from the same measured
mode amplitudes $\mathcal{G}_{\kk}$ and detuning $\delta$. The two
sides of Eq.~\eqref{eq:relation} are therefore measured on the same
data.

\label{sec:kerr-results}
\begin{table}[tb]
\caption{Comparison of the NIT and Koopman finite-window method at the $3\!:\!2$ crossing, for the
resonant mode $\kk = (3,-2)$. The
detuning $\delta = \kk\cdot\Ups$ is measured from the orbital fundamental
frequencies; $|w_{\kk}| = |\mathcal{G}_{\kk}/\delta|$ is the near-identity
coefficient; $|\chi_{\kk}|$ is the finite-window coefficient,  and the last two rows are its analytic bound
$|\mathcal{G}_{\kk}|\Lambda$ and their ratio. }
\label{tab:boundedness}
\begin{ruledtabular}
\begin{tabular}{lccc}
 & Eq. & Far              & Crossing \\
  &      & $p = 6.40M$ & $p = 5.60M$ \\
\colrule
$\delta$                              & \eqref{eq:detuning} & $3.3\cdot10^{-1}$ & $9.7\cdot10^{-5}$ \\
$|w_{\kk}|$ (near-identity)           & \eqref{eq:nit}      & $3.15\cdot10^{-5}$ & $8.59\cdot10^{-2}$ \\
$|\chi_{\kk}|$ (finite window)        & \eqref{eq:magnus}   & $3.06\cdot10^{-5}$ & $2.80\cdot10^{-5}$ \\
bound $|\mathcal{G}_{\kk}|\Lambda$    & \eqref{eq:bound}    & $3.19\cdot10^{-5}$ & $2.80\cdot10^{-5}$ \\
$|\chi_{\kk}| / |\mathcal{G}_{\kk}|\Lambda$ & \eqref{eq:bound} & $0.958$ & $1$ \\
\end{tabular}
\end{ruledtabular}
\end{table}


Recall from Sec.~\ref{sec:koopman} that the near-identity
coefficient $w_{\kk}$ applies the operator
$(1-U_\Lambda)^{-1}$, which diverges wherever an eigenvalue
$e^{\mathrm{i}\delta\Lambda}$ reaches unity, so on approach to a
crossing it grows as $|\mathcal{G}_{\kk}/\delta|$ without bound. On the other hand, the
finite-window coefficient $\chi_{\kk}$ of Eq.~\eqref{eq:magnus}
applies the logarithm, which remains finite there and obeys
$|\chi_{\kk}| \le |\mathcal{G}_{\kk}|\Lambda$ at every detuning
[Eq.~\eqref{eq:bound}]. In what follows  we evaluate both on the same
measured spectrum; the accuracy of the method far from a crossing is
quantified on the post-Newtonian benchmark in App.~\ref{app:pn}.\\

\begin{figure}[!t]
\includegraphics[width=\columnwidth]{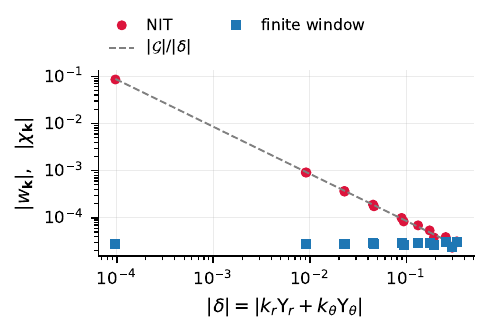}
\caption{The NIT and Koopman  coefficients against the detuning $|\delta|$ ,
showing the $|\mathcal{G}|/|\delta|$ law (dashed, no free parameter)
followed by the near-identity coefficient over four decades and \emph{not}
followed by the finite-window one, which is flat. The near-identity
curve is the operator $(1-U_\Lambda)^{-1}$ built from the same
$U_\Lambda$. Its growth is what makes locating the crossing
necessary.}
\label{fig:boundedness-detuning}
\end{figure}

At the crossing, where $\delta\Lambda \ll 1$, Eq.~\eqref{eq:relation}
reduces to $\chi_{\kk} \simeq \mathcal{G}_{\kk}\Lambda$, while the
near-identity coefficient grows as $|\mathcal{G}_{\kk}/\delta|$
without bound, so the two coefficients separate there.
Table~\ref{tab:boundedness} and Fig.~\ref{fig:boundedness-crossing}
show this separation for the resonant mode $\kk = (3,-2)$, the
$3\!:\!2$ crossing. Between the far point at $p = 6.40M$ and the
crossing the near-identity coefficient is amplified by three orders
of magnitude, and it follows the $|\mathcal{G}|/|\delta|$ law over
four decades in the detuning (Fig.~\ref{fig:boundedness-detuning}).
The finite-window coefficient changes by less than $10\%$ over the
same scan, and it attains the bound of Eq.~\eqref{eq:bound} to better
than one part in $10^{6}$ (Fig.~\ref{fig:boundedness-bound}). Because
the detuning changes sign across the scanned range, from
$+3.3\cdot10^{-1}$ to $-3.0\cdot10^{-1}$, the crossing is a zero of
$\delta(p)$ independently of how we located it. We also verify that
the numerical quadrature of Eq.~\eqref{eq:magnus} agrees with the
closed form to $6.0\cdot10^{-17}$, and that the agreement is exact at
$\delta = 0$.

\subsection{Recovery of the resonant jump by a finite window}
\label{sec:kick}

\begin{figure}[!t]
\includegraphics[width=\columnwidth]{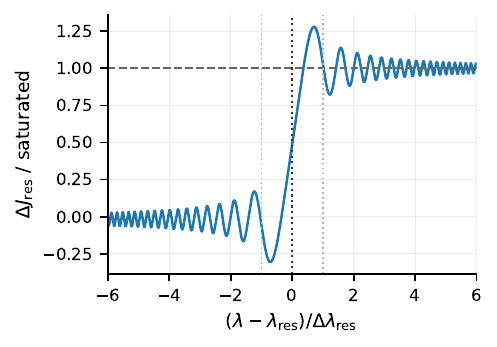}
\caption{The transient crossing. The running resonant integral is
shown against Mino time in units of the crossing width of
Eq.~\eqref{eq:transient}, normalised to its saturated value.}
\label{fig:kick-integral}
\end{figure}

At the $3\!:\!2$ crossing the resonant mode is $\kres = (3,-2)$, and
its detuning is swept through zero as the inspiral advances.
Linearising the detuning across the crossing,
$\delta(\lambda) \simeq \dot\delta\,(\lambda - \lambda_{\rm res})$,
we find that the resonant phase $q_\perp = \kres\cdot\bm q$, whose
rate of advance is the detuning of Eq.~\eqref{eq:detuning},
accumulates quadratically,
$q_\perp \simeq q_0 + \tfrac12\dot\delta(\lambda-\lambda_{\rm res})^2$,
where $q_0$ is its value at the crossing. Because the mode
contributes coherently only while $q_\perp$ stays within about half
an oscillation of its stationary value, setting the accumulated
phase equal to $\pi$ gives the crossing duration
\begin{equation}
  \Delta\lambda_{\rm res} \;\sim\; \sqrt{2\pi/|\dot\delta|} .
  \label{eq:transient}
\end{equation}
The running resonant integral, normalised to its saturated value,
acquires essentially all of the jump within one width of the
crossing and rings down towards unity from both sides
(Fig.~\ref{fig:kick-integral}).

\begin{figure}[!t]
\includegraphics[width=\columnwidth]{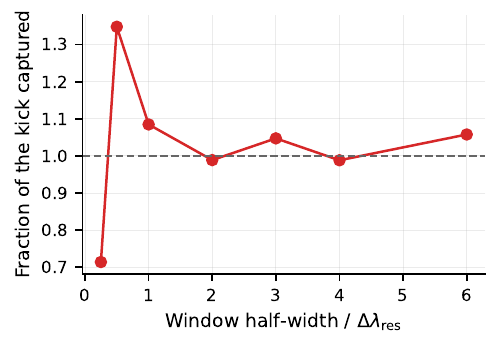}
\caption{Fraction of the total jump captured by a window of
half-width $n$ crossing widths [Eq.~\eqref{eq:transient}]. The
overshoot to $1.34$ at half a width is the first Fresnel oscillation
visible in Fig.~\ref{fig:kick-integral}; from two widths onwards the
captured fraction stays within a few percent of unity.}
\label{fig:kick-window}
\end{figure}

\begin{figure}[!t]
\includegraphics[width=\columnwidth]{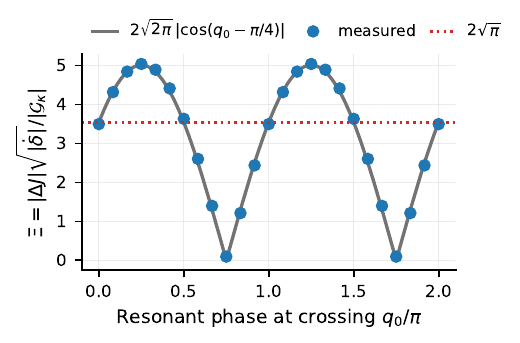}
\caption{The dependence of the jump on the resonant phase at
crossing. Each point is the jump recovered from a full sweep through
the crossing at that value of $q_0$ (Sec.~\ref{sec:kick}); the solid
curve is the envelope
$2\sqrt{2\pi}|\cos(q_0-\pi/4)|$ of Eq.~\eqref{eq:kick-phase}, and the
dotted line marks $2\sqrt{\pi}$, the value at $q_0 = 0$; the two differ by
$\sqrt2$. The predicted zeros at $q_0 = 3\pi/4$ and $7\pi/4$ are reproduced.}
\label{fig:kick-phase}
\end{figure}

\paragraph*{Saturation of the jump.} A finite window must be wide
enough to contain the jump. To measure how wide, we truncate the
running integral of Fig.~\ref{fig:kick-integral} at $\pm n$ crossing
widths about the crossing. Dividing the truncated value by the
saturated one gives the fraction of the jump that a window of that
half-width captures (Fig.~\ref{fig:kick-window}). A quarter width
captures $0.709$ of the jump. Half a width overshoots, at $1.34$. One
width gives $1.08$, and $n = 2$, $3$, $4$ give $0.982$, $1.04$ and
$0.983$. The fraction does not approach unity from one side. It rings
about it, due to the Fresnel integral of
Eq.~\eqref{eq:kick-sum} and its tail contributes in alternating
half-oscillations. The envelope of the ringing decays as the inverse
distance from the crossing in units of the width of
Eq.~\eqref{eq:transient}. A window of two crossing widths therefore
already contains the jump to a few percent.

\paragraph*{Coefficient of the resonant jump.} To quantify  the resonance coefficient  we define the ratio
\begin{equation}
  \Xi \;\equiv\;
  \frac{|\Delta J_{\rm res}|\,\sqrt{|\dot\delta|}}{|\mathcal{G}_{\kres}|} .
  \label{eq:xi}
\end{equation}

\begin{figure}[!t]
\includegraphics[width=\columnwidth]{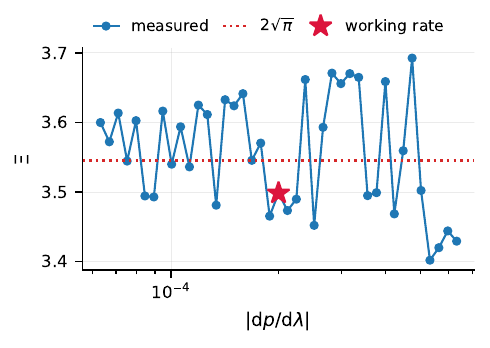}
\caption{The coefficient $\Xi$ against the sweep
rate. The residual variation is the endpoint
systematic of App.~\ref{app:verification}.}
\label{fig:kick-rates}
\end{figure}

Figure~\ref{fig:kick-rates} shows the coefficient $\Xi$ of
Eq.~\eqref{eq:xi} against the sweep rate $|\dd p/\dd\lambda|$, over a
factor of $10$ around the rate
$\dd p/\dd\lambda = -2\cdot10^{-4}$ used throughout this section
(star), with $2\sqrt\pi$ dotted. $\Xi$ stays within $3.40$--$3.69$
while the raw jump varies by $\sqrt{10}$, so the flatness of this
curve carries the $|\dot\delta|^{-1/2}$ scaling of the jump and its
level carries the stationary-phase coefficient.

We observe that the recovered jump depends, at the few-percent level,
on where the sweep is cut. The quoted $\pm 0.09$ is this finite-window
systematic, not an accuracy. We therefore report $\Xi$ as the mean and
standard deviation over half-ranges from $4$ to $10.5$ crossing widths
of an extended grid (App.~\ref{app:verification}). The systematic is
common to numerator and denominator, so no ratio-type result is
affected.

At zero resonant phase, evaluating the stationary-phase result of
Eq.~\eqref{eq:kick-phase} below at $q_0 = 0$ and inserting it into
Eq.~\eqref{eq:xi} predicts $\Xi = 2\sqrt{\pi} = 3.54$. We measure $\Xi = 3.52 \pm 0.09$, a
ratio of $0.993 \pm 0.026$.\footnote{The sweep shown in
Figs.~\ref{fig:kick-integral}--\ref{fig:kick-phase} is sampled at $468$
points per radian of resonant phase, so it is resolved by a wide margin.}
We see that by varying the sweep rate by a factor of $10$ 
the raw jump moves by $\sqrt{10}$
with a fitted exponent of $-0.5$ to within $0.03$, while $\Xi$ stays
within $3.40$--$3.69$.\\

\paragraph*{Transient-resonance jumps and phase dependence.} In what
follows we study whether the \emph{finite-window} construction
reproduces the transient-resonance jumps, which are functions of the
resonant phase $q_0$ at the crossing. Applying stationary phase to
$\int \mathcal{G}_{\kres} e^{\mathrm{i}q_\perp(\lambda)}\dd\lambda$
with $q_\perp \simeq q_0 +
\tfrac12\dot\delta(\lambda-\lambda_{\rm res})^2$, we obtain one
contribution per stationary point, $s = \pm1$,

\begin{equation}
  \Delta J_{\rm res} = \sum_{s=\pm1}
    \sqrt{\frac{2\pi}{|\dot\delta|}}\,
    \exp\!\Big[\mathrm{i}\,\mathrm{sgn}(\dot\delta s)\frac{\pi}{4}
               + \mathrm{i} s q_0\Big]\,\mathcal{G}_{s\kres} ,
  \label{eq:kick-sum}
\end{equation}
which is Eq.~(2.11) of Ref.~\cite{GuptaEtAl2022}. For a real force the
Fourier coefficients obey
$\mathcal{G}_{-\kres} = \overline{\mathcal{G}_{\kres}}$, so the two
stationary-point contributions are complex conjugates of one another and
their sum collapses to
\begin{equation}
  |\Delta J_{\rm res}| = 2|\mathcal{G}_{\kres}|
    \sqrt{\frac{2\pi}{|\dot\delta|}}\;
    \Big|\cos\Big(q_0 + \frac{\pi}{4}\,\mathrm{sgn}\,\dot\delta\Big)\Big| .
  \label{eq:kick-phase}
\end{equation}

Equations~\eqref{eq:kick-sum} and~\eqref{eq:kick-phase} are the
transient-resonance jumps of Ref.~\cite{FlanaganHinderer2012}, and
Fig.~\ref{fig:kick-phase} shows that the agreement is $0.96$--$0.99$
across $q_0 \in [0,2\pi)$. At $q_0 = 3\pi/4$ the cosine vanishes, the
measured $\Xi$ falls to $7.98\cdot10^{-2}$ against an envelope of
$5.01$, and only the subleading term survives. Notice that a model with the wrong
phase dependence would not vanish at the predicted phase.

\subsection{Behaviour between crossings}
\label{sec:offres}
In this section we study the two coefficients away from resonance
crossings, where an inspiral spends most of its time and errors can
accumulate. Dividing Eq.~\eqref{eq:relation} by the near-identity
coefficient we obtain
\begin{equation}
  \frac{|\chi_{\kk}|}{|w_{\kk}|}
  \;=\; \big|e^{\mathrm{i}\delta\Lambda}-1\big|
  \;=\; 2\big|\sin(\delta\Lambda/2)\big|.
  \label{eq:offres-ratio}
\end{equation}

\begin{figure}[!t]
\includegraphics[width=\columnwidth]{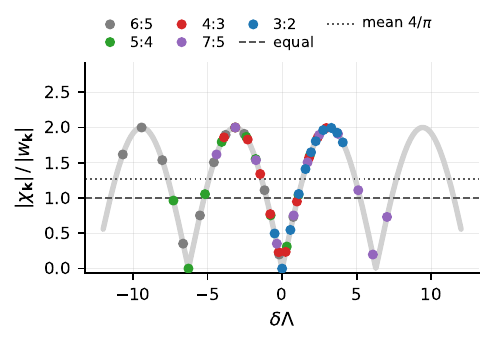}
\caption{The ratio $|\chi_{\kk}|/|w_{\kk}|$ away from a crossing,
against $\delta\Lambda$. Colour gives the
mode, and the five low-order modes are shown at each of the eleven measured
radii, with $2|\sin(\delta\Lambda/2)|$ drawn in grey. The ratio equals unity
only at isolated detunings, and averages $4/\pi$.}
\label{fig:offres-collapse}
\end{figure}

\begin{figure}[!t]
\includegraphics[width=\columnwidth]{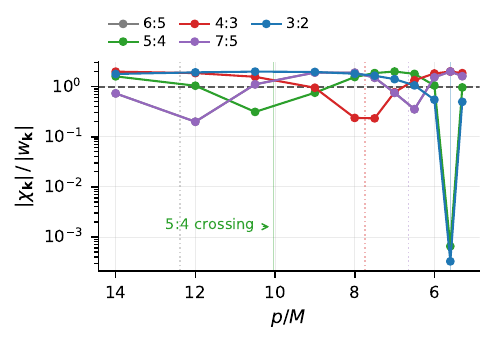}
\caption{The ratio $|\chi_{\kk}|/|w_{\kk}|$ against $p$ on a log
scale, with each mode's crossing marked, solid for the two the parity
rule of App.~\ref{sec:silent-mode} allows and dotted for the three it
forbids. The ratio vanishes at $\delta\Lambda = 2\pi n$ as
well as at $\delta = 0$, so the one dip visible in the scanned range is a
window blind to that mode rather than a crossing; four of the five
crossings fall between samples and leave no trace.}
\label{fig:offres-inspiral}
\end{figure}

Notice that this is a function of $\delta\Lambda$ alone, consequently no
amplitude, no mode index, and no property of the background enters
it. This is shown in Fig.~\ref{fig:offres-collapse}, where all
measured points fall on the single curve $2|\sin(\delta\Lambda/2)|$.
Off resonance the ratio oscillates in $[0,2]$ and averages
$4/\pi = 1.27$, so the two coefficients are of the same order between
crossings. The ratio also vanishes at $\delta\Lambda = 2\pi n$, where
the window is blind to a physically present mode because the
eigenvalue $e^{\mathrm{i}\delta\Lambda}$ of the Koopman operator
$U_\Lambda$ has returned to unity without a resonance; in
Fig.~\ref{fig:offres-inspiral} such a zero of the sinc function and a
crossing are indistinguishable by eye.

We repeat the same study with a \emph{tidal} resonance in place of
the drag-driven one.\label{sec:tidal} In this case the resonance
condition involves all three fundamental frequencies,
$n\Omega_r + k\Omega_\theta + m\Omega_\phi =
0$~\cite{GuptaEtAl2021,GuptaEtAl2022}. By axisymmetry the $m = 0$
modes are degenerate with self-force resonances, and in what follows
we set $(n,k,m) = (-3,2,0)$.


\begin{table}[t]
\caption{Measured jump for the $3\!:\!2$ crossing, mode
$\kres = (3,-2)$, driven by the drag force and by the published tidal
amplitude of Ref.~\cite{GuptaEtAl2022}, with the equation defining
each quoted magnitude.}
\label{tab:tidal}
\begin{ruledtabular}
\begin{tabular}{lccc}
 & Eq. & drag & tidal fit \\
\colrule
$|\mathcal{G}_{\kres}|$      & \eqref{eq:forcing}    & $8.32\cdot10^{-6}$ & $2.26\cdot10^{-1}$ \\
$|\Delta J_{\rm res}|$       & \eqref{eq:kick-sum}   & $3.05\cdot10^{-3}$ & $8.29\cdot10^{+1}$ \\
$\Xi$                        & \eqref{eq:xi}         & $3.52 \pm 0.09$ & $3.52 \pm 0.09$ \\
$\Xi/2\sqrt\pi$              & \eqref{eq:kick-phase} & $0.99 \pm 0.03$ & $0.99 \pm 0.03$ \\
\end{tabular}
\end{ruledtabular}
\end{table}

Table~\ref{tab:tidal} shows the measured jump for the $3\!:\!2$
crossing, mode $\kres = (3,-2)$, driven by the drag force and by the
published tidal amplitude~\cite{GuptaEtAl2022}. The mode amplitude
$|\mathcal{G}_{\kres}|$ of Eq.~\eqref{eq:forcing} is measured by the
torus projection of App.~\ref{sec:torus} (not by trajectory
averaging) for the drag force, and taken from the fit of
Ref.~\cite{GuptaEtAl2022} for the tidal one; $2\sqrt\pi$ is the value
of Eq.~\eqref{eq:kick-phase} at $q_0 = 0$. The driving differs by four orders of magnitude and $\Xi$ does not
move. The two $\Xi/2\sqrt\pi$ entries each carry the few-percent
finite-window systematic of Sec.~\ref{sec:kick}, but their
\emph{difference} does not, since the systematic is common to both
columns. We can also see that the jump scales linearly with the
forcing term---the jump of Eq.~\eqref{eq:kick-sum} is first order in
the perturbation, with $\mathcal{G}_{\kres}$ entering as an overall
factor, so rescaling the driving must rescale the jump by the same
factor and leave the normalised coefficient $\Xi$ of
Eq.~\eqref{eq:xi} unchanged. A nonlinear response of the crossing,
for instance the jump feeding back on the sweep or capture into the
resonance, would have broken this proportionality for the tidal
driving, the larger of the two amplitudes by a factor
$2.7\cdot10^{4}$.

\subsection{An estimate of the self-force amplitude}
\label{sec:selfforce}

To estimate how strongly the \emph{self-force} itself drives a
resonance we need the resonant amplitude in units of the adiabatic
rate on which it appears, $|\mathcal{G}_{\kres}|/|\langle\dot
J\rangle|$. This ratio sets the size of the resonant kick relative to
the smooth inspiral. We obtain it from the flux tables of
Ref.~\cite{FlanaganHughesRuangsri2014}. On a resonant orbit the
averaged flux depends on the resonant phase, and that dependence is
established there to be sinusoidal; the modulation is then
characterised by a single amplitude, its half peak-to-trough excursion,
which in our convention is $|\mathcal{G}_{\kres}|$ itself. Their
tabulated fractional peak-to-trough variation $\Delta X$ [their
Eq.~(4.5)] therefore converts directly,
\begin{equation}
  \frac{|\mathcal{G}_{\kres}|}{|\langle\dot J\rangle|}
  \;=\; \frac{\Delta X}{2}.
  \label{eq:fhr-amplitude}
\end{equation}
The largest variation over their sixteen resonant Kerr geodesics at
$a = 0.9M$ occurs on the $3\!:\!2$ crossing and gives
$\tfrac12\cdot 1.03\% = 5.15\cdot10^{-3}$, and no entry in their
four tables exceeds $6\cdot10^{-3}$. The gravitational self-force
therefore drives a strong-field transient resonance at about half a
percent of the adiabatic rate it rides on.

We stress that their grid does not contain
our configuration, so we quote the nearest corner rather than
interpolate. Their tables record the extremes over the resonant
phase, so the phase itself remains a free parameter. And their own
accuracy caveat places the four smallest entries near the truncation
level of the calculation.

Under the reflection $\theta \to \pi - \theta$ a driving of definite
azimuthal number $m$ can excite only modes of matching parity, which
forbids the $4\!:\!3$ crossing while allowing the $3\!:\!2$. We derive
this rule in App.~\ref{sec:silent-mode} for the drag force; the same
rule holds exactly in the tidal calculation of
Ref.~\cite{GuptaEtAl2022}, and since the two drivings share only the
background geodesics, this identifies it as a property of the
background rather than of the probe (App.~\ref{sec:parity-lit}). The
self-force, however, is a sum over $m$, and that sum contains a
matching-parity term for every mode, so no crossing is forbidden. 

The size of the residual is measured in the tables of
Ref.~\cite{FlanaganHughesRuangsri2014}---their $4\!:\!3$ variation is
$22$ to $127$ times smaller than the $3\!:\!2$, suppressed but not
zero. Under self-force driving the selection rule therefore survives
as a suppression by one to two orders of magnitude rather than as an
exact zero; the crossings the rule forbids for a driving of definite
$m$ appear as the dashed detuning curves of
Fig.~\ref{fig:sequence-detunings}.

\subsection{A sequence of crossings}
\label{sec:sequence}

\begin{figure}[!t]
\includegraphics[width=\columnwidth]{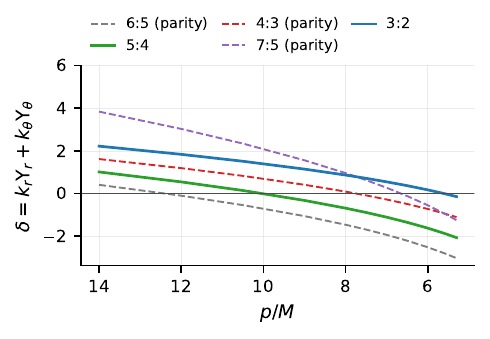}
\caption{The detuning $\delta = \kk\cdot\Ups$ of Eq.~\eqref{eq:detuning}
for each low-order mode $\kk$ in the scanned range, against the
semi-latus rectum $p$. Each curve falls monotonically through zero, so
each resonance is crossed exactly once, at the point where its curve
meets $\delta = 0$. Solid curves are the two crossings a driving
symmetric about the equatorial plane can excite; dashed curves are the
three the parity rule of App.~\ref{sec:silent-mode} forbids.}
\label{fig:sequence-detunings}
\end{figure}

\begin{figure}[!t]
\includegraphics[width=\columnwidth]{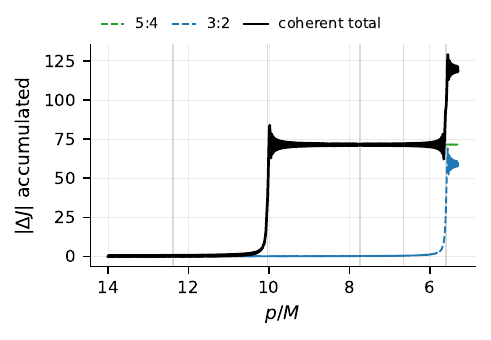}
\caption{The running resonant integral through all five crossings in
a single window, with no per-crossing bookkeeping and no branch. The
two live contributions are dashed and the coherent total solid; grey
lines mark all five locations.}
\label{fig:sequence-accumulation}
\end{figure}

\begin{figure}[!t]
\includegraphics[width=\columnwidth]{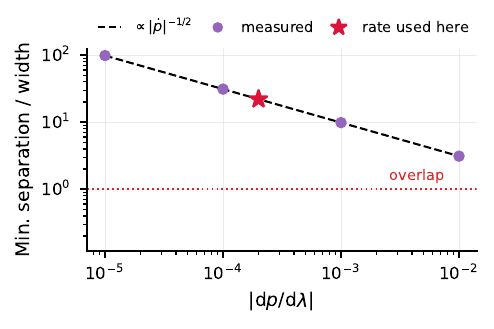}
\caption{The minimum separation between crossings in units of their width,
against sweep rate.}
\label{fig:sequence-overlap}
\end{figure}

It is expected that an inspiral sweeps through several
resonances, and their effects
accumulate~\cite{Levati2025}. We measure both fundamental
frequencies, $\Upsilon_r$ and $\Upsilon_\theta$, across $p \in [5.3, 14]M$.
Each crossing sits where the detuning $\delta = \kk\cdot\Ups$ of
Eq.~\eqref{eq:detuning} vanishes, located by bisection, and its width is the stationary-phase width
$\Delta\lambda_{\rm res} = \sqrt{2\pi/|\dot\delta|}$ of
Eq.~\eqref{eq:transient}. We find five crossings, listed in
Table~\ref{tab:sequence} and
shown in Figs.~\ref{fig:sequence-detunings}--\ref{fig:sequence-overlap}:

\begin{table}[t]
\caption{Low-order crossings in the scanned range, at
$\dd p/\dd\lambda = -2\cdot10^{-4}$. Only even $k_\theta$ can be excited by a
driving symmetric about the equatorial plane (App.~\ref{sec:silent-mode}).}
\label{tab:sequence}
\begin{ruledtabular}
\begin{tabular}{lccccc}
mode & $(k_r,k_\theta)$ & $p_{\rm res}/M$ & width & sep./width & excitable \\
\colrule
$6\!:\!5$ & $(6,-5)$ & $12.391$ & $339$ & $36$ & no \\
$5\!:\!4$ & $(5,-4)$ & $10.036$ & $321$ & $37$ & \textbf{yes} \\
$4\!:\!3$ & $(4,-3)$ & $7.742$  & $297$ & $22$ & no \\
$7\!:\!5$ & $(7,-5)$ & $6.652$  & $198$ & $23$ & no \\
$3\!:\!2$ & $(3,-2)$ & $5.600$  & $262$ & --- & \textbf{yes} \\
\end{tabular}
\end{ruledtabular}
\end{table}

In the last column of Table~\ref{tab:sequence}, a crossing is marked
excitable when its $k_\theta$ is even, so that a driving symmetric
about the equatorial plane places nonzero amplitude on its resonant
mode (App.~\ref{sec:silent-mode}); a crossing marked no is swept
through but receives no kick. Three consequences follow:

\paragraph{Effect of the parity rule.} The first result is the satisfaction of
the selection rule. Three of the five crossings have odd $k_\theta$ and
are forbidden; only the $5\!:\!4$ and the $3\!:\!2$ survive. The parity
rule of App.~\ref{sec:silent-mode} is therefore not a statement about one
crossing but a property of the whole sequence.

\paragraph{Isolation of the crossings.} The second result is that the
crossings stay separate. Since
$\dot\delta = (\dd\delta/\dd p)(\dd p/\dd\lambda)$, the width of a
crossing scales as $|\dd p/\dd\lambda|^{-1/2}$ while the gap between
crossings scales as $|\dd p/\dd\lambda|^{-1}$, so
\begin{equation}
  \frac{\rm separation}{\rm width} \;\propto\;
  \Big|\frac{\dd p}{\dd\lambda}\Big|^{-1/2} .
  \label{eq:overlap}
\end{equation}
This is opposite to what one might expect: a slow sweep remains longer at each crossing, but the width grows only as the square root while the gap grows linearly, so crossings \emph{separate} as the sweep slows. At
$\dd p/\dd\lambda = -2\cdot10^{-4}$ we measure a separation of
$22$--$37$ widths, where for each adjacent pair of crossings the gap
in Mino time between their locations is divided by the mean of their
widths (Table~\ref{tab:sequence}); the ratio reaches unity only near
$\dd p/\dd\lambda = -10^{-1}$, a rate some $500$ times faster
(Fig.~\ref{fig:sequence-overlap}). Overlapping resonances would
invalidate the treatment of each crossing as an isolated
stationary-phase event; this is the standard criterion of
Chirikov~\cite{Chirikov1979}, confirmed in the case of the single-window construction integrating across all five crossings at once. Fig.~\ref{fig:sequence-accumulation} shows that at any physically relevant rate each crossing is a separate event.

\paragraph{Composition of the resonance kicks.} The third result
concerns how the two surviving kicks combine. We integrate the
response through the whole range in a single pass, so one window
covers both crossings and neither requires individual treatment, see Fig.~\ref{fig:sequence-accumulation}. Each surviving crossing
contributes a kick $\Delta J_k$, a complex number whose modulus is
bounded by the envelope
$2|\mathcal{G}_{\bm k}|\sqrt{2\pi/|\dot\delta|}$ and whose phase is
set by the resonant phase at that crossing; we find that the measured kicks sit at
$0.492$ and $0.495$ of their envelopes, with their fractions fixed by each
crossing's phase [Eq.~\eqref{eq:kick-phase}]. Because the kicks arrive with different phases, the total is the
complex sum of the two, and its magnitude is approximately the sum of
the magnitudes: $|\sum_k \Delta J_k| / \sum_k|\Delta J_k| = 0.925$\footnote{The
value $0.925$ is an extrapolation in the sample count $n$, which it
approaches as $1/n$: we measure $0.912$, $0.919$ and $0.922$ at
$n = 4, 8, 16\cdot10^{5}$, with successive differences halving to
within $1\%$.}. It is not exactly 1 due to the partial cancellation of
the relative arrival phase.

How much of each kick survives is set by the resonant phase on
arrival, $q_0 = \int\delta\,\dd\lambda$, the detuning of
Eq.~\eqref{eq:detuning} accumulated from the start of the sweep to the
crossing. Despite this phase is predictable in principle, in practice, however, we find that at the
$3\!:\!2$ it reaches $5.3\cdot10^{4}$ radians, so placing $q_0$ within one
radian would require $\delta(p)$ to a relative accuracy of
$1.9\cdot10^{-5}$, beyond the accuracy expected for a frequency model, see
e.g.\ Refs.~\cite{SperiGair2021,Levati2025}.

\subsection{Caveats and limitations}
\label{sec:kerr-caveats}
\label{sec:vs-partial}
\label{sec:limits}
\label{sec:sf-gap}

This study comes with some caveats and limitations.
\begin{enumerate}
\item Neither driving is a gravitational self-force, since no
resonant self-force amplitude is published at absolute
normalisation~\cite{FlanaganHughesRuangsri2014}. The sweep rate is
likewise derived from a different force than the driving, and
corresponds to a mass ratio $\eps \approx 6\cdot10^{-5}$. Repeating
the crossing at the derived rates gives $\Xi/2\sqrt\pi = 0.985$ at
$\eps = 10^{-5}$ and $0.970$ at $\eps = 10^{-4}$, and both values
sit inside the systematic of Sec.~\ref{sec:kick}. The tidal fits of
Sec.~\ref{sec:tidal} hold $\mathcal{G}_{\kres}$ fixed across the
crossing; the drag force varies by $20\%$ over the same range, and
holding it fixed changes $\Xi$ by $0.10\%$.

\item The off-resonance fraction and the off-resonance accuracy come
from different systems. We measure the fraction on Kerr, where the
sequence of Sec.~\ref{sec:sequence} spends $98.7\%$ of the scanned
range outside the crossing widths of Table~\ref{tab:sequence}, and
the accuracy on a planar binary at $\nu = 1/4$ (App.~\ref{app:pn}). For this reason we cannot provide a single net figure.

\item The comparison with partial
averaging~\cite{vandeMeent2014,Lynch2024} is structural rather than
numerical: partial averaging covers the crossing with a second
transformation and a tuned switching criterion, whereas we propose a
single operator in place of a branch at a tuned threshold. The
advantage of Eq.~\eqref{eq:relation} is confined to detunings below
the window resolution $2\pi/\Lambda$, since off resonance the
windowed generator oscillates with amplitude
$2|\mathcal{G}_{\kk}|/|\kk\cdot\Ups|$, which is the size of the
denominator it replaces. Whether the structural advantage is also a
numerical one is a question of cost at matched accuracy, which will be studied in a future work. The optimally tuned switch of
Ref.~\cite{Lynch2024} attains a phase error of $O(\eps^{4/7})$, and
the corresponding scaling for the finite window at a crossing has
not been measured, so the two schemes cannot yet be ranked.
\end{enumerate}

%% file: sections/summary.tex
\section{Summary and discussion}
\label{sec:summary}
\label{sec:discussion}

In this work we have presented a new approach to transient resonance
crossings based on the Koopman
operator~\cite{Koopman1931,Mezic2005}. The evolution over one radial
cycle is encoded in the operator $U_\Lambda$, and the averaging
generator is built from its logarithm as an entire function of the
operator. As a consequence, no resonance need be located in
advance, no second transformation constructed, and no switching
criterion invoked.

We have tested our approach on a Kerr geodesic driven through a
 $3\!:\!2$ transient resonance. On the orbit where the
near-identity coefficient is amplified by $2.7\cdot10^{3}$, the
finite-window generator attains the analytic bound of
Eq.~\eqref{eq:bound} to better than one part in $10^{6}$. A window of
two crossing widths then recovers the leading resonant jump, and we
measure $\Xi = 3.52 \pm 0.09$ against the stationary-phase prediction
$2\sqrt\pi = 3.54$. The predicted dependence on the crossing phase is
reproduced as well, including its zero~\cite{FlanaganHinderer2012}.

We find that the resonant jump tracks the
tidal amplitudes of Refs.~\cite{GuptaEtAl2021,GuptaEtAl2022}, and the first-order generator matches the
closed forms of Ref.~\cite{Blanco2026} to $10^{-15}$. Beyond the
single crossing, we have integrated a sequence of five crossings in
one window and verified their isolation and coherent composition
(Sec.~\ref{sec:sequence}). We have also measured the off-resonance
cost against orbit averaging on a planar post-Newtonian benchmark
(App.~\ref{app:pn}) and reduced the second Magnus coefficient
$\chi_{(2)}$ to an integral on the physical phase space
(App.~\ref{app:chi2}).

We note that the near-identity coefficient is the operator $(1-U_\Lambda)^{-1}$
evaluated next to a spectral degeneracy, whereas the finite-window
coefficient is an integral of an entire integrand over a compact
interval, at every Magnus order (Sec.~\ref{sec:all-orders}). This leads to the following obsrvations:  i) there is nothing to locate, since $\delta$
enters only through $\sinc(\delta\Lambda/2)$, ii) there is nothing to
switch, since one generator is applied uniformly, and iii) there is
nothing to regularise, since no denominator is ever formed. 

To benchmarks our approach, we have calibrated the fundamental frequencies and their uncertainty against the sixteen resonant geodesics of
Ref.~\cite{FlanaganHughesRuangsri2014} (App.~\ref{sec:external}) .

Finally, writing the inspiral as a linear operator acting on phase-space
observables moves the modelling problem from the trajectory to the
operator. This  setting opens the door to modern numerical linear
algebra techniques that are fast growing~\cite{MartinssonTropp2020, Higham2008,GazzolaSabateLandman2020}. In this framework the object to
be approximated or compressed is a single map rather than a branch
structure. Such a map can be amortised, since it can be learned once as a
function of the system parameters and evaluated cheaply thereafter. Structure-preserving architectures for this already exist for the
conservative sector, such as the symplectic neural flows of
Ref.~\cite{Canizares2024}. This suggests a division of work by
structure rather than by timescale, with a representation
for the symplectic part of the dynamics and a generator
representation for the dissipative part. This division would only be efficient only where
the one-cycle average is itself expensive, as for 
the self-force-driven inspiral.

%% file: sections/appendices.tex
\section{The Koopman operator under radiation reaction}
\label{app:koopman}

The unitarity of $U_t$ follows from measure preservation, and
radiation reaction is not measure preserving. The actions drift, the
phase-space volume in the action--angle variables $(\bm{q},J)$ is not
conserved, and the flow has no invariant measure on any bounded region
containing the inspiral. Writing the eigenvalues of the one-cycle map
as $e^{\varpi_j\Lambda}$, the exponents are purely imaginary for a
measure-preserving flow, $\varpi_j = \mathrm{i}\,\kk\cdot\Ups$ as in
Eq.~\eqref{eq:koopman}, and acquire a real part under dissipation.
However the generator
$\chi = \log U_\Lambda$ of Eq.~\eqref{eq:onecycle} is defined from the
dissipative one-cycle map directly and not from a spectral
decomposition, so it is not affected by the breakdown of unitarity. In fact, what the real part bounds is the regime in which the
unitary spectral picture of Sec.~\ref{sec:observable} may be quoted, we quantify this below.

To measure the exponents, samples of a set of observables separated by
one radial cycle are fitted onto one another, and
$\varpi_j = \log\mu_j/\Lambda$ is obtained from the eigenvalues $\mu_j$ of
the fitted matrix. For conservative (geodesic) motion the real parts
must vanish, so their measured size validates any benchmark estimator we use in one
case and measures the dissipation in the other.

Normalising each real part by its own frequency
$\omega_j = |\mathrm{Im}\,\varpi_j|$,
\begin{align}
  |\mathrm{Re}\,\varpi|/\omega
    &\lesssim 1\cdot10^{-9}
    &&\text{(reaction off)},\label{eq:growth-off}\\
  |\mathrm{Re}\,\varpi|/\omega
    &\simeq 2.5\text{--}4.0\cdot10^{-2}
    &&\text{(reaction on)}.\label{eq:growth-on}
\end{align}
Equation~\eqref{eq:growth-off} confirms unitarity nine orders below the
measured frequencies; Eq.~\eqref{eq:growth-on} shows the spectrum
leaving the unit circle at the few-percent level; equivalently, for
the fundamental mode, whose phase advances by $\omega\Lambda = 2\pi$
per radial cycle, the modulus of an observable changes by a factor
$e^{2\pi\,|\mathrm{Re}\,\varpi|/\omega} \simeq 1.17$--$1.29$ over one
window. Since the inspiral is a transient, there is no single
stationary value to quote, and the range given is the spread obtained by
repeating the propagator fit above over different windows along the
evolution.


\subsection{A perturbation carries no amplitude at the resonant mode}
\label{sec:silent-mode}

One consequence of Eq.~\eqref{eq:bound} concerns drivings that carry
no amplitude at the resonant index. If $\mathcal{G}_{\kres} = 0$, the
resonance is present in the geodesics but the perturbation cannot
excite it, so a crossing experiment measures no jump. The null is a
property of the chosen driving; however, it reads as a failure of the
construction. Two physically natural drivings fall in this class: the
gravitational-wave fluxes of a circular orbit, and an unmodulated
drag force.

The first is given by $\dot L_z = \dot E\cos\iota/\Omega_\phi$ with $\dot Q = 2Q\dot L_z/L_z$.
This driving term is a function of the orbit-averaged elements $(E,L_z,Q,\bar r)$ alone (we measure an inclination drift of
$\sim 10^{-7}$ over a full inspiral), and a function of the actions alone carries only the $\kk = \bm 0$ Fourier mode of
Eq.~\eqref{eq:forcing}, that is, its amplitude at every resonant index is zero, so it has zero contribution through a crossing.

Similarly, consider a drag force of the form
$f_a = -\alpha[1 + \beta\cos(m\theta)]\,p_a$, where $a$ is a
spacetime index, $p_a$ the covariant four-momentum, $\theta$ the
Boyer--Lindquist polar angle, $\alpha$ sets the strength of the drag,
and $\beta$ and $m$ the amplitude and harmonic of its polar
modulation. Its driving is
\begin{equation}
  \mathcal{G} = -\alpha E\big[r(q_r)^2 + a^2\cos^2\theta(q_\theta)\big]
                \big[1 + \beta\cos m\theta\big].
  \label{eq:drag}
\end{equation}
For $k_r \neq 0$ only the $r^2$ term survives, so the coefficient factorises
as $\mathcal{G}_{k_r k_\theta} = [r^2]_{k_r}\times[1+\beta\cos m\theta]_{k_\theta}$.
At $\beta = 0$ the second factor is the constant $1$, whose only
Fourier coefficient is $k_\theta = 0$. Every mode with
$k_\theta \neq 0$ is therefore \emph{exactly} zero, independently of
the angular dependence.

Notice that the same factorisation leads to a parity selection
rule. Under $\theta \to \pi - \theta$, even $m$ feeds even $k_\theta$
and odd $m$ feeds odd. The $4\!:\!3$ crossing, of mode $(4,-3)$
(Table~\ref{tab:sequence}), needs $k_\theta = -3$, and
$\cos\theta(q_\theta)$ is nearly sinusoidal, so its third harmonic is
four orders of magnitude down ($5.7\cdot10^{-10}$ against
$9.9\cdot10^{-6}$). \emph{The $4\!:\!3$ crossing therefore does not
contribute}, and the usable configuration is the $3\!:\!2$ crossing,
mode $(3,-2)$, driven by an even modulation $m = 2$.

We accordingly adopt $\beta = 0.5$, $m = 2$, $\alpha = 10^{-3}$, and verify by
direct projection that $\mathcal{G}_{(3,-2)} \neq 0$.

The same selection rule was noticed in Ref.~\cite{GuptaEtAl2022}, where the
authors compute tidal-resonance jumps by a different route
and report that the jumps \emph{vanish} whenever $k+m$ is odd. Their
zero is exact, enforced by the symmetry of the tidal potential, where
ours is a four-order suppression instead; since the two drivings share only
the background geodesics, the agreement identifies the rule as a
property of the background rather than of the probe.

\section{Numerical methods}
\label{app:numerics}
\label{sec:implementation}

\subsection{Extracting the mode amplitude: projection onto the torus}
\label{sec:torus}

An adiabatic evolution depends only on the secular term
$\mathcal{G}_{\bm 0}$. However, the leading resonant jump is set by the amplitude
$\mathcal{G}_{\kres}$ of the harmonic that stops oscillating at the
crossing. The amplitude is a coefficient of the two-torus Fourier expansion of the driving, whose
$\kk = \bm 0$ term is the average
\begin{equation}
  \langle f\rangle_{q_r,q_\theta}
  \;=\; \frac{1}{(2\pi)^2}\!\int_0^{2\pi}\!\!\!\int_0^{2\pi}
        f(q_r,q_\theta)\,\dd q_r\,\dd q_\theta .
  \label{eq:torus-average}
\end{equation}
The standard adiabatic treatment obtains these coefficients by averaging along the trajectory, which is valid only when the orbit
fills the torus densely (see Ref.~\cite{SperiGair2021}). This approach is therefore not valid on a resonant torus where the
orbit closes after finitely many cycles and covers a one-dimensional
curve rather than the full $(q_r, q_\theta)$ torus, and the
trajectory average fails, returning a spurious value even for a
driving with no resonant content.

To see this, we build a $256\times256$ tensor-product grid in $(q_r,q_\theta)$ (in Mino time the radial and polar motions decouple
exactly~\cite{Mino2003,Schmidt2002}, so the driving is a proper
function on a two-torus) and take a two-dimensional discrete
Fourier transform; its coefficients are the mode amplitudes
$\mathcal{G}_{\kk}$, of which Eq.~\eqref{eq:torus-average} is the
$\kk = \bm{0}$ one. We find that for the above unmodulated drag ($\beta = 0$) every mode
with $k_\theta \neq 0$ vanishes identically. The trajectory average
on the resonant torus at $p = 6.2M$ nevertheless returns
$|\mathcal{G}_{(3,-2)}| = 4.0\cdot10^{-5}$, a leakage of the same
order as the true amplitudes of the crossing, while the projection
returns $\sim\!10^{-20}$ on the same mode. For this reason we extract every
amplitude by projecting the orbit onto the invariant torus, since averaging there requires no ergodicity assumption, involves no window length, and does not degrade near a resonance.

\subsection{The finite-window systematic on $\Xi$}
\label{app:verification}
\label{app:xi-systematic}
\label{sec:silent-object}

The $\pm 0.09$ quoted on $\Xi$ in Sec.~\ref{sec:kick} is a
finite-window systematic, not a statistical error. We find that $\Xi$
converges to six significant figures by $1.25\cdot10^{4}$ samples, and
that the measured inputs, $\delta(p)$ and $\mathcal{G}(p)$
(App.~\ref{app:kerr-validation} and App.~\ref{sec:torus}), are
reproduced at the $10^{-4}$ level; the remainder is the endpoint of
the window, whose Fresnel ringing leaves $4$--$5\%$ between integer
numbers of crossing widths. Concretely, over an extended grid spanning
$p \in [5.05, 6.15]M$, that is $\pm10.5$ crossing widths, the value of
$\Xi$ recovered varies between $0.963$ and $1.05$ times the prediction
$2\sqrt\pi$ depending on where the window is cut; the quoted
$\pm 0.09$ is the standard deviation of that variation.

\section{Kerr frequency validation}
\label{app:kerr-validation}
\label{sec:external}

\begin{figure}[!t]
\includegraphics[width=\columnwidth]{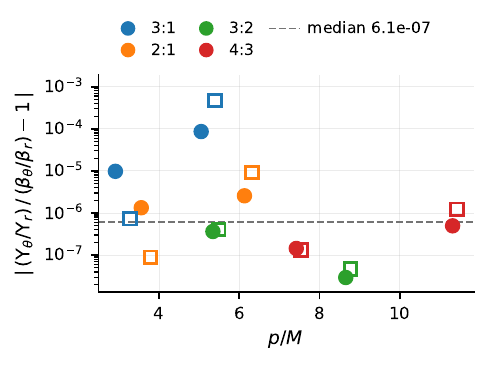}
\caption{Calibration of our computed fundamental frequencies
$\Upsilon_\theta$ and $\Upsilon_r$ against the sixteen resonant
geodesics of Ref.~\cite{FlanaganHughesRuangsri2014}. The dashed line
is the median relative error, $6.1\cdot10^{-7}$.}
\label{fig:calibration-freq}
\end{figure}

\begin{figure}[!t]
\includegraphics[width=\columnwidth]{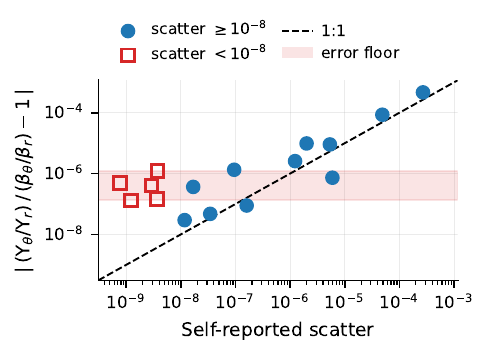}
\caption{Calibration of the error bar itself, on the same sixteen
orbits of Ref.~\cite{FlanaganHughesRuangsri2014}. On the dashed
$1\!:\!1$ line the self-reported scatter equals the true error; below
a scatter of $10^{-8}$ the true error saturates and the self-report is
no longer informative.}
\label{fig:calibration-errorbar}
\end{figure}

The frequencies $\Upsilon_\theta$ and $\Upsilon_r$ used to compute the
detuning, Eq.~\eqref{eq:detuning}, are validated against values in
the literature. Evaluating $\Upsilon_\theta/\Upsilon_r$ at the
sixteen resonant Kerr geodesics of Ref.~\cite{FlanaganHughesRuangsri2014} at
$a = 0.9M$ we reproduce all sixteen commensurabilities, with a median relative
error of $6.1\cdot10^{-7}$ and a worst case of $4.7\cdot10^{-4}$ in the
strong field (Fig.~\ref{fig:calibration-freq}). Because the true error is
thereby known, we use this procedure to calibrate our own uncertainty. Figure~\ref{fig:calibration-errorbar} shows that the two track one
another down to a self-reported scatter of $10^{-8}$. Below that
threshold the true error stops improving and saturates at a floor
near $10^{-7}$, so the self-reported figure is no longer informative
there. The threshold and the floor are different quantities. The
first is read on the self-report axis and the second on the
true-error axis.

\section{The post-Newtonian benchmark}
\label{app:pn}

In general, resonance crossings are rare events, and the orbital
evolution is off resonance for most of the inspiral.\label{sec:pn}
For that reason we study the planar system, which carries no resonance
at any order, as a null test for the off-resonance accuracy, since it
quantifies any error accumulated away from a crossing.

That the planar system carries no resonance follows from a rotational
symmetry rather than from any numerical
obstacle.\label{sec:nulls}\label{sec:no-resonance} The Delaunay
angles are the mean anomaly $\ell$ and the argument of periastron
$g$. The periastron advances at first post-Newtonian order by
$6\pi M/p$ per radial cycle~\cite{DamourDeruelle1985}, so the two
frequencies are independent. The obstruction to a resonance lies
elsewhere. The angle $g$ orients the orbit within its own plane, and
a planar system has no preferred direction there. Neither the
conservative Hamiltonian nor the reaction force can therefore depend
on $g$. Every mode has $k_g = 0$, and the detuning reduces to
$k_\ell\,\Upsilon^{\ell}$. That combination vanishes only for the
secular mode, which a near-identity transformation excludes by
construction.

\begin{figure}[!t]
\includegraphics[width=\columnwidth]{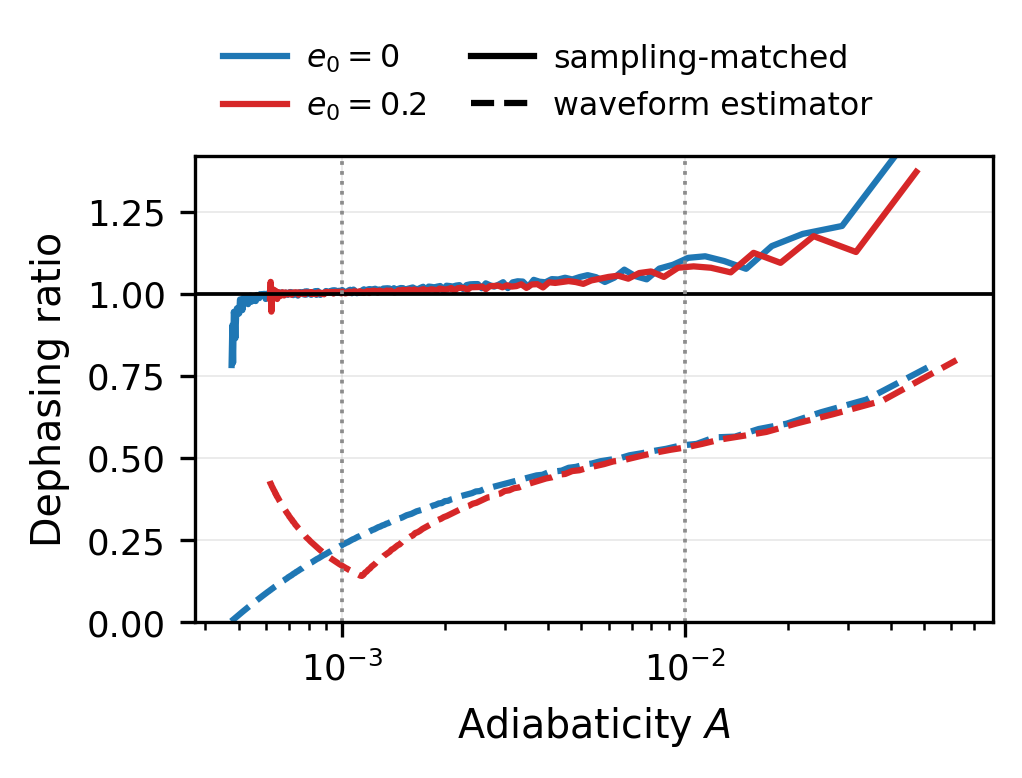}
\caption{Ratio of the orbit-averaged to the flow-map dephasing
against the adiabaticity parameter $A = |\dot\Omega_r|/\Omega_r^2$,
at $e_0 = 0$
and $0.2$ and $\nu = 1/4$, on the two estimators: sampling-matched at
cycle boundaries (solid) and the dense-grid waveform estimator
(dashed).}
\label{fig:breakdown}
\end{figure}

We use the
reaction force in the harmonic-gauge form of Ref.~\cite{Blanco2026},
Eq.~(4.2), and evolve the system in three different ways: (i) a
resolved reference integration of the equations of motion, (ii)
standard orbit averaging, and (iii) the windowed flow map. Each starts
from $r_0 = 40M$ at $\nu = 1/4$ and runs into the regime where we
expect orbit averaging itself to degrade (Fig.~\ref{fig:breakdown}).

For the flow-map route the generator is calibrated first. We compute
$\chi_{(1)}$ both from the closed form of Ref.~\cite{Blanco2026},
Eq.~(4.21), and by direct quadrature from the raw reaction force. The
two routes reproduce three of the four Delaunay components to
$10^{-15}$, including the null $\chi_{\mathcal{G}} = 0$.\footnote{The fourth
component agrees once the quartic coefficient printed as $36$ is read
as $37$, the value required for consistency with Peters' eccentricity
enhancement function~\cite{Peters1964}.} The closed forms and the
quadrature share only the force term, so their agreement validates
everything built on top of it. As a validation for the
evolution itself we use the \code{EccentricTD} and \code{TaylorT4}
waveform models, generated through LALSuite~\cite{LALSuite}. 

Notice that the flow-map approach outputs one sample per radial
cycle, whereas orbit averaging can be output densely. If the
one-per-cycle output is interpolated onto a dense grid before
comparison, the interpolation sets an error floor of its own. For this
reason we report two estimators: (a) the dense-grid \emph{waveform
estimator}, which measures the phase model each route delivers to a
downstream user, interpolation included, and (b) the
\emph{sampling-matched} estimator, which compares the averaging
approximations themselves at the cycle boundaries where both routes
produce a value. Figure~\ref{fig:breakdown} shows the ratio of the
orbit-averaged to the flow-map dephasing on both estimators, at
$e_0 = 0$ and $0.2$, against the adiabaticity parameter
$A \equiv |\dot\Omega_r|/\Omega_r^2$, the fractional change of the
radial frequency $\Omega_r$ per radian of orbital phase, with the dot
here a coordinate-time derivative; adiabatic evolution means
$A \ll 1$, and $A$ grows as the orbit shrinks, with $A = 10^{-3}$ and
$10^{-2}$ reached at $r/M = 29.7$ and $11.8$ for these runs (vertical
lines). We find that the sampling-matched ratio crosses unity near
$A \approx 10^{-3}$ and rises to $1.42$ and $1.37$ at the deep end,
while on the waveform estimator the ratio stays below unity, which
reflects the interpolation of the one-per-cycle output rather than the
averaging scheme. The first cycles of the sampling-matched curves are
start-up transients, where the accumulated dephasing is still near
zero and the ratio is ill-conditioned.\\
\break

Sampled alike, the flow-map route is the more accurate of the two at
every mass ratio measured. On the waveform estimator the
interpolation penalty competes with this gain, so there the flow map
overtakes orbit averaging only below a crossover mass ratio, which two
independent determinations bracket between $\eps = 1.6\cdot10^{-6}$
and $3.3\cdot10^{-6}$. The crossover marks the point where the
sampling penalty, which is independent of $\eps$, stops outweighing
the flow map's extra power of $\eps$ in accuracy.

\section{Reduction of the second Magnus coefficient to a physical-space integral}
\label{app:chi2}

Section~\ref{sec:all-orders} bounds every Magnus coefficient of a
finite window; here we carry the second coefficient to a form in which
it can be evaluated. For a dissipative force $\chi_{(2)}$ is defined
on the Galley-doubled space~\cite{Galley2013,Tsang2015}, where
Ref.~\cite{Blanco2026} gives it as a double time-ordered integral of a
doubled bracket of interaction Hamiltonians. The time-ordered integral
collapses by bilinearity of the bracket,
\begin{widetext}
\begin{equation}
  \int_0^\Lambda\!\!\dd s_1\!\int_0^{s_1}\!\!\dd s_2\,
  \big[\tilde F(s_1), \tilde F(s_2)\big]
  = \int_0^\Lambda\!\!\dd s_1\,
    \Big[\tilde F(s_1),\, \textstyle\int_0^{s_1} \tilde F(s_2)\,\dd s_2\Big],
  \label{eq:chi2-collapse}
\end{equation}
\end{widetext}
the inner integral being the running primitive of the pullback of
Eq.~\eqref{eq:pullback}, accumulated alongside the outer pass. The
cost is one pass over the cycle, the same as for $\chi_{(1)}$, plus
the derivative of the pulled-back field with respect to the reference
point; boundedness in the detuning is the $n = 2$ case of
Eq.~\eqref{eq:simplex}, already established.

We do not evaluate $\chi_{(2)}$ for the Kerr system here and Eq.~\eqref{eq:chi2-collapse} is given as a reduction and not yet as a
validated estimator.